\documentclass[
aps,pra,reprint,numerical,amsmath,
amssymb,amsfonts,showkeys,showpacs]{revtex4-1}
\usepackage{bm,graphicx,xcolor,hyperref,rotating,lineno,multirow}
\newcommand{\mc}{\multicolumn}
\newcommand{\mr}{\multirow}
\newcommand{\md}{\mathcal{D}}

\begin{document}

\title{Excited states of spherium}

\author{Pierre-Fran\c{c}ois Loos}
\email{loos@rsc.anu.edu.au}
\author{Peter M. W. Gill}
\thanks{Corresponding author}
\email{peter.gill@anu.edu.au}
\affiliation{Research School of Chemistry, Australian National University, Canberra, ACT 0200, Australia}
\date{\today}

\begin{abstract}
We report analytic solutions of a recently discovered quasi-exactly solvable model consisting of two electrons, interacting {\em via} a Coulomb potential, but restricted to remain on the surface of a $\md$-dimensional sphere.  Polynomial solutions are found for the ground state, 
and for some higher ($L\le3$) states.  Kato cusp conditions and interdimensional degeneracies are discussed.
\end{abstract}

\keywords{
Exact solution, excited states, spherium, cusp condition, interdimensional degeneracies}

\maketitle

\section{
\label{sec:intro}
Introduction}

A quasi-exactly solvable model is one for which it is possible to solve the Schr\"odinger equation exactly for a finite portion of the energy spectrum \cite{Ushveridze}.  In quantum chemistry, a famous example of this is the Hooke's law atom \cite{KestnerPhysRev1962, WhiteJCP1970, Kais89, TautPRA1993}, which consists of a pair of electrons, repelling Coulombically but trapped in a harmonic external potential.  This model and others \cite{AlaviJCP2000, JungJCP2003, JungPRA2004, ThompsonPRB2002, ThompsonPRB2004, ThompsonJPCM2004, ThompsonJCP2005} have been used extensively to test various approximations \cite{FilippiJCP1994, TautJPB1998, IvanovJCP1999, SeidlPRL2000, SeidlPRA2007b, SunJCTC2009, GoriGiorgiIJQC2009, SeidlPRA2010} within density functional theory (DFT) \cite{HohenbergPRB1964, KohnPRA1965, ParrYang} and explicitly correlated methods \cite{KutzelniggTheorChemAcc1985, KutzelniggJChemPhys1991, HendersonCPL2001, HendersonPRA2004, BokhanPCCP2008}.

We have recently discovered \cite{Quasi09} that a pair electrons, repelling Coulombically but constrained to remain on the surface of a $\md$-sphere of radius $R$ \cite{EzraPRA1982, EzraPRA1983, OjhaPRA1987, HindePRA1990, SeidlPRA2007b, TEOAS1, TEOAS2, EcLimit09, Loos10} is also quasi-exactly solvable and we have called this system $\md$-spherium.  (We adopt the convention that a $\md$-sphere 
is the surface of a ($\md+1$)-dimensional ball.)  We have shown that the Schr\"odinger equation for the $^1S$ and the $^3P$ states of $\md$-spherium can be solved exactly for a countably infinite set of $R$ values and that the resulting wave functions are polynomials in the interelectronic distance $0 \le u \equiv |\bm{r}_1-\bm{r}_2| \le 2R$.

In this article, we extend our earlier results \cite{Quasi09} to higher angular momentum (up to $L=3$) states of $\md$-spherium ($\md \ge 2$)
for both the singlet and triplet manifolds.  The $\md = 1$ case is anomalous and, for brevity, is not discussed here.  We use atomic units throughout.

\section{
\label{sec:wave}
Wave function}

The Hamiltonian of $\md$-spherium is
\begin{equation}
\label{H}
	\hat{H} =  
	- \frac{1}{2} \left( \nabla_1^2 + \nabla_2^2 \right)
	+ \frac{1}{u},
\end{equation}
where the two first terms represent 
the kinetic contribution of each electron,
and $u^{-1}$ is the Coulomb operator.

\begin{table*}
\caption{
\label{tab:summary}
Ground state and excited states of $\md$-spherium}
\begin{ruledtabular}
\begin{tabular}{cccccccc}
State	&	Configuration		&	$\chi (\bm{\Omega}_1,\bm{\Omega}_2)$					&	$\delta$	&	$\gamma^{-1}$		&	$\Lambda$	& $\kappa$		&	Transformation	\\
\hline                                                                                                                                                                                                                                                          
$^1S^{\rm e}$	&	$s^2$		&	1									&	$2\md-1$	&	$\md-1$		&	0		&	0	&	$^3P^{\rm e}$	\\
\hline
$^3P^{\rm o}$	&	$sp$		&	$\cos \theta_1 - \cos \theta_2$						&	$2\md+1$       	&	$\md+1$		&	$\md/2$		&	1	&	$^1D^{\rm o}$	\\
$^1P^{\rm o}$	&	$sp$		&	$\cos \theta_1 + \cos \theta_2$						&	$2\md+1$	&       $\md-1$		&	$\md/2$		&	0	&	$^3D^{\rm o}$	\\
$^3P^{\rm e}$	&	$p^2$		&	$\sin \theta_1 \sin \theta_2 \sin \left( \phi_1 - \phi_2 \right)$	&	$2\md+3$	&	$\md+1$		&	$\md$		&	1	\\
\hline
$^3D^{\rm e}$	&	$sd$		&	$\chi_{^3P^{\rm o}} \cdot \chi_{^1P^{\rm o}}$					&	$2\md+3$	&	$\md+1$		&	$\md+1$		&	1	&$^1F^{\rm e}$	\\
$^1D^{\rm o}$	&	$pd$		&	$\chi_{^3P^{\rm o}} \cdot \chi_{^3P^{\rm e}}$					&	$2\md+5$	&	$\md+3$		&	$3\md/2+1$	&	2	\\
$^3D^{\rm o}$	&	$pd$		&	$\chi_{^1P^{\rm o}} \cdot \chi_{^3P^{\rm e}}$					&	$2\md+5$	&	$\md+1$		&	$3\md/2+1$	&	1	\\
\hline
$^1F^{\rm e}$ 	&	$pf$		&	$\chi_{^3P^{\rm e}} \cdot \chi_{^3D^{\rm e}}$					&	$2\md+7$	&	$\md+3$		&	$2\md+3$	&	2	\\
\end{tabular}
\end{ruledtabular}
\end{table*}

Following Breit \cite{BreitPR1930}, 
we write the total wave function as the product
\begin{equation}
\label{Phi}
	\Phi (\{s_1,s_2\},\{\bm{\Omega}_1,\bm{\Omega}_2\},u)
	= \Xi(s_1,s_2) \chi (\bm{\Omega}_1,\bm{\Omega}_2) \Psi(u),
\end{equation}
where $\Xi$, $\chi$ and $\Psi$ are the spin, 
angular and interparticle wave functions, respectively, and
$s_i$ and $\bm{\Omega}_i$ are the spin and hyperspherical coordinates 
\cite{Louck60} of the $i$-th electron.
The singlet and triplet wave functions are given by the familiar \cite{BetheSalpeter} forms
\begin{align}
	^1\Xi(s_1,s_2) & 
	= \frac{1}{\sqrt{2}} \left[ \alpha(s_1) \beta(s_2) - \beta(s_1) \alpha(s_2) \right],		\\
	^3\Xi(s_1,s_2) & = 
	\begin{cases}
		\alpha(s_1) \alpha(s_2),								\\
		\frac{1}{\sqrt{2}} \left[ \alpha(s_1) \beta(s_2) + \beta(s_1) \alpha(s_2) \right],	\\
		\beta(s_1) \beta(s_2).
	\end{cases}
\end{align}
The angular part is associated with an energy 
\begin{gather}
	E_{\chi} = \frac{\Lambda}{R^2}, 	\\
	\Lambda = \frac{\ell_1(\ell_1+\md-1)}{2} 
	+ \frac{\ell_2(\ell_2+\md-1)}{2},
\end{gather}
where $\ell_1$ and $\ell_2$ are the angular momentum
quantum numbers of the corresponding one-electron 
configuration ($s=0$, $p=1$, $d=2$, $f=3$, \ldots).
The functions $\chi$, 
which are dependent on the nature of the state considered 
\cite{BreitPR1930,KingJCP67}, 
are gathered in Table \ref{tab:summary}, 
where $\theta_i \in [0,\pi]$ and $\phi_i \in [0,2\pi]$ 
are the $(\md-1)$-th and $\md$-th 
hyperspherical angles of the electron $i$.
The corresponding one-electron configurations
are also reported.
In Table \ref{tab:summary}, 
the suffixes e (even) and o (odd) 
are related to the parity of the states, 
which is given by $(-1)^{\ell_1+\ell_2}$.
Hence, we label the states with the notation 
$^{1,3}L^{\rm e,o}$, where $L=S,P,D,F,\ldots$

\section{
\label{sec:poly}
Polynomial solutions}

Substituting the ansatz \eqref{Phi} 
into the Hamiltonian \eqref{H} yields
the Schr\"odinger equation
\begin{equation}
\label{H-u}
 	\left( \frac{u^2}{4R^2} - 1 \right) \frac{d^2\Psi}{du^2} 
	+ \left(\frac{\delta u}{4R^2} - \frac{1}{\gamma u} \right)\frac{d\Psi}{du} 
	+ \frac{\Psi}{u} = E \Psi,
\end{equation}
where the parameters $\delta$ and $\gamma$ are tabulated for each state in Table \ref{tab:summary}.

By introducing the dimensionless variable $x = u/2R$, 
Eq. \eqref{H-u} can be recast as
a Heun's differential equation \cite{Ronveaux} 
with singular points at $x = -1, 0, +1$.
Following the known solutions of 
this equation \cite{Polyanin}, 
we seek wave functions of the form
\begin{equation} 
\label{series}
	\Psi(u) = \sum_{k=0}^\infty a_k\,u^k,
\end{equation}
and substitution into \eqref{H-u} yields 
the three-step recurrence relation
\begin{multline}
	a_{k+2} = \frac{\gamma}{(k+2)\left[(k+1)\gamma+1\right]} 
	\bigg\{ a_{k+1} \\
	+ \left[\frac{k(k+\delta-1)}{4R^2} - E\right] a_{k} \bigg\},
\end{multline}
with the starting values $a_0 = 1$ and $a_1 = \gamma$.

If the series \eqref{series} terminates
at a certain $k = n$, we obtain the exact wave function
\begin{equation}
	\Psi_{n,m}(u) 
	= \sum_{k=0}^{n} a_{k}\,u^{k},
\end{equation}
for a particular radius 
$R_{n,m}$ and energy $E_{n,m}$.
This is an $n$th degree polynomial with $m$ nodes 
between $0$ and $2R$ 
($ 0 \leq m \leq \lfloor \frac{n+1}{2} \rfloor$)
and requires that $a_{n+1}$ and $a_{n+2}$ vanish.  
If $a_{n+1} = 0$, the relation 
\begin{equation}
\label{Enm}
	R_{n,m}^2 E_{n,m} = \frac{n}{2}\left(\frac{n}{2}+\frac{\delta-1}{2}\right) 
\end{equation}
ensures that $a_{n+2} = 0$.
For a given $n$, the energies are thus determined 
by finding the roots of the equation $a_{n+1} = 0$, 
which is a polynomial in $E$, 
of degree $\lfloor \frac{n+1}{2} \rfloor$.

For the $^1D^{\rm e}$ state, 
we have not been able to obtain
polynomial solutions because 
the Hamiltonian \eqref{H} is not 
separable using the ansatz \eqref{Phi}
and the wave function satisfies
exchange coupled equations \cite{HerrickJMP75}.
This applies also to some other states of higher 
angular momentum.

\section{
\label{sec:res}
Results and Discussion}

\begin{table*}
\caption{
\label{tab:TEOAS-1Po}
Radii $R_{n,m}$ and energies $E_{n,m}$ for $^1P^{\rm o}$ states 
of two electrons on a $\md$-sphere ($\md$=2,3,4)}
\begin{ruledtabular}
\begin{tabular}{cccccccccccccc}
&		&	\mc{4}{c}{$\md = 2$}					&	\mc{4}{c}{$\md = 3$}					&	\mc{4}{c}{$\md = 4$}					\\
			\cline{3-6}							\cline{7-10}							\cline{11-14}
&	$n$/$m$	&	0	&	1	&	2	&	3	&	0	&	1	&	2	&	3	&	0	&	1	&	2	&	3	\\
\hline
\mr{9}{*}{\rotatebox{90}{Radius}}                                                                                                                                                                                              
&	1	&1.118		&		&		&		&1.871		&		&		&		&2.598		&		&		&		\\
&	2	&3.162		&		&		&		&4.637		&		&		&		&6.083		&		&		&		\\
&	3	&6.226		&1.656		&		&		&8.376		&2.520		&		&		&10.52		&3.303		&		&		\\
&	4	&10.30		&4.232		&		&		&13.11		&5.888		&		& 		&15.93		&7.440		&		&		\\
&	5	&15.38		&7.847		&2.159		&		&18.84		&10.21		&3.127		&		&22.32		&12.49		&3.966		&		\\
&	6	&21.46		&12.49		&5.246		&		&25.57		&15.53		&7.077		& 		&29.72		&18.51		&8.732		&		\\
&	7	&28.54		&18.14		&9.397		&2.639		&33.30		&21.84		&11.97		&3.707		&38.11		&25.51		&14.39		&4.599		\\
&	8	&36.63		&24.80		&14.59		&6.222		&42.03		&29.15		&17.86		&8.220		&47.49		&33.49		&21.01		&9.976		\\
\hline
\mr{9}{*}{\rotatebox{90}{Energy}}                                                                                                                                                                                              
&	1	&1.000		&		&		&		&0.5000		&		&		&		&0.3333		&		&		&		\\
&	2	&0.3000		&		&		&		&0.1860		&		&		&		&0.1351		&		&		&		\\
&	3	&0.1355		&1.914		&		&		&0.09622	&1.063		&		&		&0.07460	&0.7562		&		&		\\
&	4	&0.07541	&0.4467		&		&		&0.05820	&0.2884		&		& 		&0.04731	&0.2168		&		&		\\
&	5	&0.04757	&0.1827		&2.414 		&		&0.03874	&0.1319		&1.406		&		&0.03260	&0.1041		&1.033		&		\\
&	6	&0.03257	&0.09620	&0.5450		&		&0.02753	&0.07468	&0.3594		& 		&0.02378	&0.06129	&0.2754		&		\\
&	7	&0.02363	&0.05851	&0.2180		&2.764 		&0.02051	&0.04771	&0.1587		&1.656 		&0.01808	&0.04035	&0.1267		&1.241 		\\
&	8	&0.01789	&0.03903	&0.1127		&0.6200		&0.01585	&0.03295	&0.08781	&0.4144		&0.01419	&0.02853	&0.07252	&0.3215		\\
\end{tabular}
\end{ruledtabular}
\end{table*}

\begin{table*}
\caption{
\label{tab:TEOAS-3Pe}
Radii $R_{n,m}$ and energies $E_{n,m}$ for $^3P^{\rm e}$ states 
of two electrons on a $\md$-sphere ($\md$=2,3,4)}
\begin{ruledtabular}
\begin{tabular}{cccccccccccccc}
&		&	\mc{4}{c}{$\md = 2$}					&	\mc{4}{c}{$\md = 3$}					&	\mc{4}{c}{$\md = 4$}					\\
			\cline{3-6}							\cline{7-10}							\cline{11-14}
&	$n$/$m$	&	0	&	1	&	2	&	3	&	0	&	1	&	2	&	3	&	0	&	1	&	2	&	3	\\
\hline
\mr{9}{*}{\rotatebox{90}{Radius}}
&	1	&	2.291	&		&		&		&	3.000	&		&		&		&	3.708	&		&		&		\\
&	2	&	5.477	&		&		&		&	6.892	&		&		&		&	8.307	&		&		&		\\
&	3	&	9.616	&	3.006	&		&		&	11.72	&	3.748	&		&		&	13.84	&	4.478	&		&		\\
&	4	&	14.73	&	6.851	&		&		&	17.52	&	8.334	&		&		&	20.32	&	9.797	&		&		\\
&	5	&	20.84	&	11.62	&	3.676	&		&	24.30	&	13.82	&	4.453	&		&	27.78	&	16.01	&	5.208	&		\\
&	6	&	27.94	&	17.35	&	8.156	&		&	32.07	&	20.26	&	9.708	&		&	36.22	&	23.15	&	11.22	&		\\
&	7	&	36.04	&	24.05	&	13.54	&	4.315	&	40.83	&	27.66	&	15.84	&	5.128	&	45.65	&	31.25	&	18.10	&	5.909	\\
&	8	&	45.13	&	31.75	&	19.86	&	9.412   &	50.58	&	36.04	&	22.90	&	11.03   &	56.07	&	40.32	&	25.89	&	12.60	\\
\hline
\mr{9}{*}{\rotatebox{90}{Energy}}
&	1	&	0.3333	&		&		&		&	0.2500	&		&		&	      	&	0.2000	&		&		&		\\
&	2	&       0.1333	&		&		&		&	0.1053	&		&		&	        &       0.08696	&		&		&		\\
&	3	&       0.07300	&	0.7472	&		&		&	0.06002	&	0.5874	&		&	        &       0.05093	&	0.4862	&		&		\\
&	4	&       0.04607	&	0.2131	&		&		&	0.03908	&	0.1728	&		&	        &       0.03390	&	0.1459	&		&		\\
&	5	&       0.03166	&	0.1019	&	1.018	&		&	0.02752	&	0.08503	&	0.8196	&	        &       0.02429	&	0.07315	&	0.6913	&		\\
&	6	&       0.02306	&	0.05983	&	0.2706	&		&	0.02042	&	0.05117	&	0.2228	&	        &       0.01829	&	0.04477	&	0.1906	&		\\
&	7	&       0.01752	&	0.03932	&	0.1242	&	1.222	&	0.01575	&	0.03432	&	0.1046	&	0.9983  &       0.01428	&	0.03046	&	0.09079	&	0.8522	\\
&	8	&       0.01375	&	0.02778	&	0.07096	&	0.3161	&	0.01251	&	0.02464	&	0.06102	&	0.2629  &       0.01145	&	0.02215	&	0.05370	&	0.2269	\\
\end{tabular}
\end{ruledtabular}
\end{table*}

Numerical values of the energies and radii
for the $^1P^{\rm o}$ and $^3P^{\rm e}$ states
are reported in Tables \ref{tab:TEOAS-1Po} 
and \ref{tab:TEOAS-3Pe}.
Tables containing results for the $^1S^{\rm e}$ 
and $^3P^{\rm o}$ states
can be found in Ref. \cite{Quasi09}.
Numerical values of the energies and radii
for the higher angular momentum states can 
be determined using the interdimensional 
degeneracies (see Sec. \ref{subsec:interdim}).

For any given state, as $n$ increases, the radius increases and the energy decreases. 
The opposite behavior is observed with respect to $m$. 
Futhermore, as $R$ (or, equivalently, $n$) increases, 
the electrons tend to localize on opposite sides of the sphere due 
to the dominance of the Coulomb interaction as the density 
decreases \cite{TEOAS1,TEOAS2}.  Such Wigner crystallization \cite{WignerPR1934}
has also been observed in other systems \cite{ThompsonPRB2004,TautPRA1993,Ball}.

\begin{figure}
\begin{center}
\includegraphics[width=0.48\textwidth]{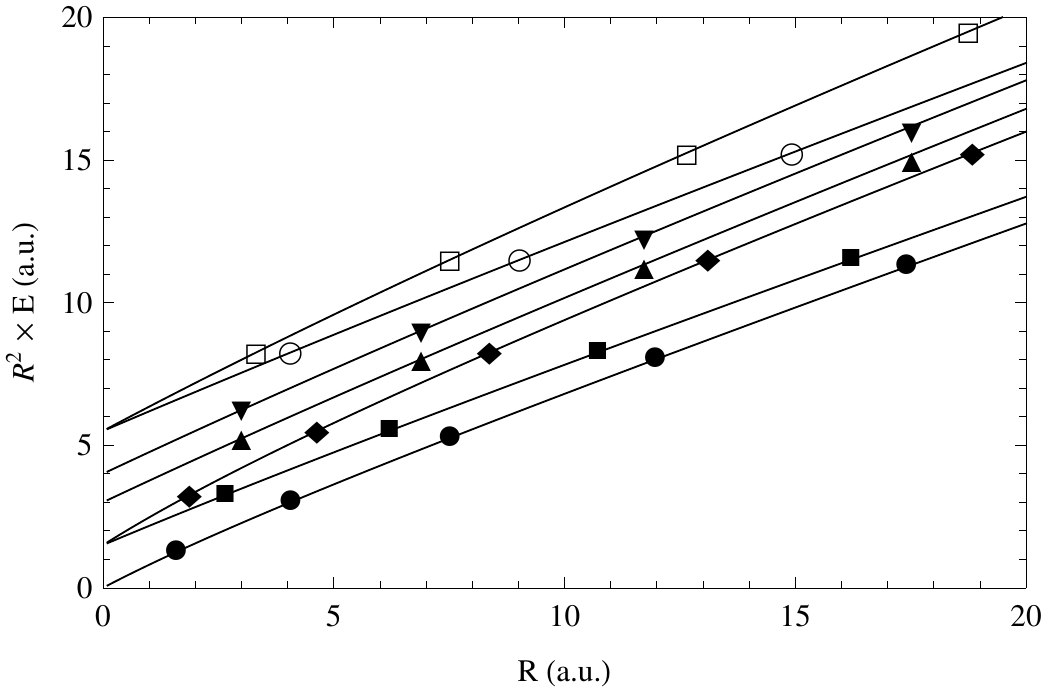}
\caption{
\label{fig:ES}
Energy of the $S$, $P$ and $D$ states of 3-spherium.
(${}^1S^{\rm e} < {}^3P^{\rm o} \leq {}^1P^{\rm o} < 
{}^3P^{\rm e} < {}^3D^{\rm e} < {}^1D^{\rm o} \leq {}^3D^{\rm o}$).
The quasi-exact solutions are shown by the markers.
}
\end{center}
\end{figure}

\begin{figure}
\begin{center}
\includegraphics[width=0.48\textwidth]{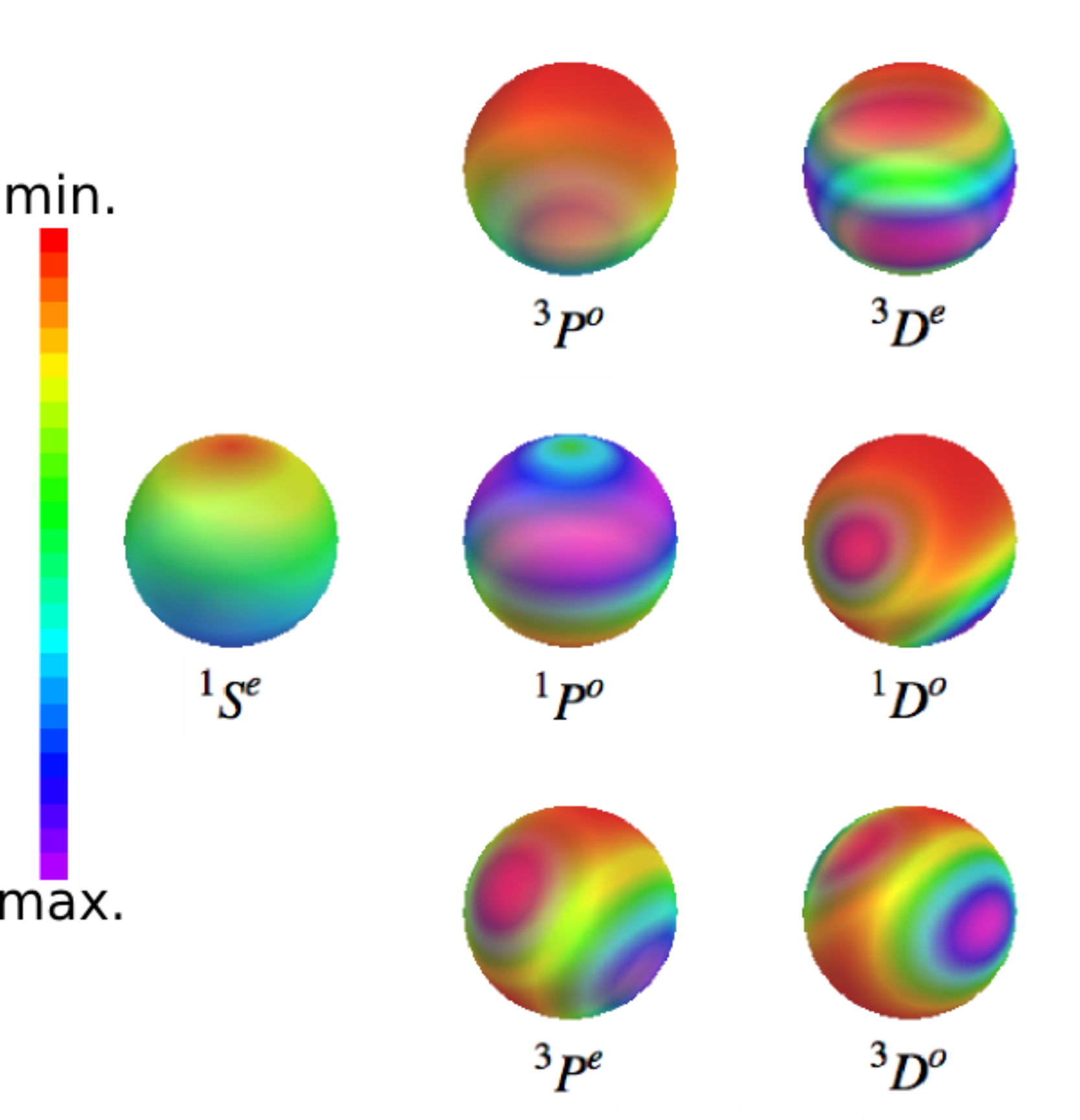}
\caption{
\label{fig:ES-on-sphere}
Density plots of the $S$, $P$ and $D$ states of 2-spherium.
The squares of the wave functions when one electron 
is fixed at the north pole are represented.
The radii are $\sqrt{3}/2$, $\sqrt{15}/2$, $\sqrt{5}/2$, 
$\sqrt{21}/2$, $\sqrt{21}/2$, $3\sqrt{5}/2$ and $3\sqrt{3}/2$
for the ${}^1S^{\rm e}$, ${}^3P^{\rm o}$, ${}^1P^{\rm o}$,
${}^3P^{\rm e}$, ${}^3D^{\rm e}$, ${}^1D^{\rm o}$
and ${}^3D^{\rm o}$ states, respectively.
}
\end{center}
\end{figure}

The energies of the $S$, $P$ and $D$ states ($m=0$)
for 3-spherium are plotted in Fig. \ref{fig:ES}
(the quasi-exact solutions are indicated by markers),
while density plots of 2-spherium ($n=1$ and $m=0$)
are represented on Fig. \ref{fig:ES-on-sphere}.

\subsection{
\label{subsec:parity}
Natural/unnatural parity}

In attempting to explain Hund's rules \cite{Hund} 
and the ``alternating'' rule \cite{Russel, Condon}
(see also \cite{BoydNature1984, WarnerNature1985}),
Morgan and Kutzelnigg 
\cite{KutzelniggJCP1992, MorganJPC1993, KutzelniggZPD1996}
have proposed that the two-electron atomic states
be classified thus:
{\em a two-electron state, composed of one-electron 
spatial orbitals with individual parities $(-1)^{\ell_1}$
and $(-1)^{\ell_1}$ and hence with overall parities 
$(-1)^{\ell_1+\ell_2}$, is said to have natural 
parity if its parity is $(-1)^L$. [\ldots]
If the parity of the two-electron state is $-(-1)^{L}$, 
the state is said to be of unnatural parity.} \cite{KutzelniggZPD1996}

After introducing spin, three classes emerge.
In a 3-dimensional space, the states with
a cusp value of 1/2 \cite{Kato1951,Kato1957}
are known as the {\em natural parity singlet states},
those with a cusp value of 1/4 \cite{PackJChemPhys1966} 
are the {\em natural and unnatural parity triplet states},
and those with a cusp value of 1/6 \cite{KutzelniggJCP1992},
are the {\em unnatural parity singlet states}.

In previous work \cite{Quasi09}, 
we have observed that the $^1S^{\rm e}$ ground state 
and the first excited $^3P^{\rm o}$ state of 3-spherium 
possess the same singlet (1/2) and triplet (1/4) cusp conditions
as those for electrons moving in three-dimensional physical space
and we have therefore argued that 3-spherium may be 
the most appropriate model for studying ``real'' 
atomic or molecular systems.
This is supported by the similarity of the 
correlation energy $E_{\rm c}$ of 3-spherium 
to that in other two-electron systems.
Indeed, it can be shown \cite{EcLimit09} that, 
as $R \to 0$, $E_{\rm c}$ approaches $-0.0476$, 
which is close to the corresponding values 
for the helium-like ions ($-0.0467$) \cite{BakerPRA1990}, 
the Hooke's law atom ($-0.0497$) \cite{GillJCP2005}, 
and two electrons in a ball ($-0.0552$) \cite{Ball}.

Most of the higher angular momentum states of 3-spherium, 
possess the ``normal'' cusp values of 1/2 and 1/4.
However, the unnatural $^1D^{\rm o}$ and $^1F^{\rm e}$ states
have the cusp value of 1/6.

\subsection{
\label{subsec:Kato-1}
First-order cusp condition}

The wave function, radius and energy of the lowest states are given by
\begin{align}
	\Psi_{1,0} (u) & = 1 + \gamma u,	&
	R_{1,0}^2 & = \frac{\delta}{4\gamma}, 	&
	E_{1,0} & = \gamma, 
\end{align}
which are closely related to the Kato cusp condition \cite{Kato1957}
\begin{equation} \label{Kato}
        \frac{\Psi^{\prime}(0)}{\Psi(0)} = \gamma.
\end{equation}

We now generalize the Morgan--Kutzelnigg classification
\cite{MorganJPC1993} to a $\md$-dimensional space.
Writing the interparticle wave function as
\begin{equation}
\label{Psi-1}
	\Psi(u) = 1 + \frac{u}{2\kappa+\md-1} + O(u^2),
\end{equation}
we have 
\begin{equation}
\label{classification}
\begin{split}
	\kappa = 0	& 	\text{ for natural parity singlet states,}	\\
	\kappa = 1	& 	\text{ for triplet states,}	\\
	\kappa = 2	& 	\text{ for unnatural parity singlet states.}
\end{split}
\end{equation}
The labels of the $\md$-spherium states
are given in Table \ref{tab:summary}.

\subsection{
\label{subsec:Kato-2}
Second-order cusp condition}

The second solution is associated with 
\begin{gather}
	\Psi_{2,0} (u) = \Psi_{1,0} (u) 
	+ \frac{\gamma ^2 (\delta +2)}{2 \gamma  (\delta +2)+4 \delta +6} u^2,	\\
	R_{2,0}^2 = \frac{(\gamma+2)(\delta+2)-1}{2\gamma},			\\
	E_{2,0} = \frac{\gamma(\delta+1)}{(\gamma+2)(\delta+2)-1}.
\end{gather}
For $\md$-spherium, the second-order cusp condition is
\begin{equation}
\label{Psi2}
	\frac{\Psi^{\prime\prime}(0)}{\Psi(0)} 
	= \frac{1}{2\md} \left( \frac{1}{\md-1} - E \right).
\end{equation}
Following \eqref{Psi2}, the classification \eqref{classification} can be extended 
to the second-order coalescence condition, where the 
wave function (correct up to second-order in $u$) is
\begin{multline}
	\Psi(u) 
	= 1 + \frac{u}{2\kappa+\md-1} \\
	+ \frac{u^2}{2(2\kappa+\md)} \left( \frac{1}{2\kappa+\md-1} - E \right) + O(u^3).
\end{multline}
Thus, we have, for $\md = 3$,
\begin{equation}
	\frac{\Psi^{\prime\prime}(0)}{\Psi(0)} =
	\begin{cases}
	\frac{1}{6} \left( \frac{1}{2} - E \right),	&	\text{ for } \kappa = 0,	\\
	\frac{1}{10} \left( \frac{1}{4} - E \right),	&	\text{ for } \kappa = 1,	\\
	\frac{1}{14} \left( \frac{1}{6} - E \right),	&	\text{ for } \kappa = 2.
	\end{cases}
\end{equation}
For the natural parity singlet states ($\kappa=0$),
the second-order cusp condition of 3-spherium 
is precisely the second-order coalescence condition 
derived by Tew \cite{TewJCP2008}, reiterating
that 3-spherium is an appropriate model for normal
physical systems.

\subsection{
\label{subsec:Kato-3}
Third-order cusp condition}

The third-order cusp condition of 3-spherium is
\begin{equation} \label{Kato-3}
	\frac{\Psi^{\prime\prime\prime}(0)}{\Psi(0)} 
	= \frac{1}{18} 
	\left( \frac{1}{8} - E + \frac{15}{16 R^2} \right),
\end{equation}
which is similar, but not strictly equivalent,
to the one derived by Tew \cite{TewJCP2008}, 
due to the $R$-dependence of \eqref{Kato-3}.
The generalization to $\md$ dimensions is straightforward.

\subsection{
\label{subsec:interdim}
Interdimensional degeneracies}

As shown in Table \ref{tab:summary}, 
many states of $\md$-spherium have the same 
Hamiltonian \eqref{H-u} as lower angular momentum states of 
$(\md+2)$-spherium.

Using the transformation 
$(\md,L) \rightarrow (\md+2,L-1)$ 
(see Table \ref{tab:summary}),
one can see that the Hamiltonian 
of the $^3P^{\rm e}$, $^1D^{\rm o}$, 
$^3D^{\rm o}$ and $^1F^{\rm e}$ states 
for a given value of $\md$ are respectively 
identical to those for $^1S^{\rm e}$, $^3P^{\rm o}$, $^1P^{\rm o}$, 
and $^3D^{\rm e}$ states at $\md+2$.
The transformation
$(\md,L) \rightarrow (\md+2,L-1)$, 
preserves the parity of the states, but
``flips'' the spin configuration, thereby
increasing by one unit the value of $\kappa$.
In $\md$-spherium, we note that
the Hamiltonians of the $^3P^{\rm e}$ and 
$^3D^{\rm e}$ states are identical.

Similar interdimensional degeneracies, 
first noticed by van Vleck \cite{vanVleck}, 
have been observed for various systems
\cite{HerrickJMP75,HerrickPRA75,Doren86,Goodson91,DunnPRA1999}.

\section{
\label{sec:ccl}
Conclusion}

In this article, we have reported exact solutions of a 
Coulomb correlation problem, consisting of two electrons 
on a $\md$-dimensional sphere.
The Coulomb problem can be solved exactly for an infinite set 
of values of the radius $R$ for both the ground and excited states, 
on both the singlet and triplet manifolds.
The corresponding exact solutions are polynomials
in the interelectronic distance $u$.

The cusp conditions (up to third-order 
in the interelectronic distance), 
which are related to the behavior 
of the wave function at the electron-electron 
coalescence point,
have been analyzed and classified according to the 
natural or unnatural parity of the state considered.

Finally, we have shown seen that, as in other one-, two- or 
three-electron systems,
there exist interdimensional degeneracies between 
some of the states of $\md$-spherium.

\begin{acknowledgments}
PMWG thanks the NCI National Facility 
for a generous grant of supercomputer time 
and the Australian Research Council 
(Grant DP0984806) for funding.
\end{acknowledgments}


\end{document}